\documentclass{ifacconf}  

\usepackage{amsmath} 
\usepackage{amssymb}  
\usepackage{enumerate}

\usepackage{tikz}
\usetikzlibrary{arrows}
\usepackage{natbib}   
\usepackage{algorithmic}
\usepackage{algorithm}
\usepackage{savesym}
\usepackage{units}
\savesymbol{AND}
\savesymbol{OR}
\savesymbol{NOT}
\savesymbol{TO}
\savesymbol{COMMENT}
\savesymbol{BODY}
\savesymbol{IF}
\savesymbol{ELSE}
\savesymbol{ELSIF}
\savesymbol{FOR}
\savesymbol{WHILE}

\newtheorem{remark}{Remark}
\newtheorem{definition}{Definition}

\newtheorem{proposition}{Proposition}

\newcommand{\norm}[1]{\left\lVert#1\right\rVert}

\DeclareMathOperator*{\argmin}{arg\,min}

\newcommand{\diag}{\mathrm{diag}}
\newcommand{\col}{\mathrm{col}}

\newcommand{\G}{\mathcal{G}}
\newcommand{\E}{\mathcal{E}}
\newcommand{\V}{\mathcal{V}}
\newcommand{\M}{\mathcal{M}}
\newcommand{\Ni}{\mathcal{N}_i}

\newcommand\copyrighttext{%
  \footnotesize \centering
  \noindent 
\textcopyright\hspace{-0.1cm} 2020 the authors. 
\textit{This work has been accepted to IFAC (MTNS 2020) for publication under a Creative Commons Licence CC-BY-NC-ND}.
   }
\newcommand\copyrightnotice{%
\begin{tikzpicture}[remember picture,overlay]
\node[anchor=south,yshift=10pt] at (9,-13.3) {{\parbox{\dimexpr\textwidth-\fboxsep-\fboxrule\relax}{\copyrighttext}}};
\end{tikzpicture}%
}

\begin{document}
\begin{frontmatter}

	\title{Distributed Safe Learning using an Invariance-based Safety Framework\thanksref{footnoteinfo}}

	\thanks[footnoteinfo]{The research of A. Carron was supported by the Swiss National Centre of Competence in Research NCCR Digital Fabrication. The research of J. Sieber was supported by the ETH Career Seed Grant 19-18-2.}

	\author[First]{Andrea Carron}
	\author[First]{Jerome Sieber}
	\author[First]{Melanie N. Zeilinger}

	\address[First]{Institute for Dynamic Systems and Control, ETH Zurich, CH-8092 Zurich, Switzerland (e-mail: carrona@ethz.ch; jsieber@ethz.ch; mzeilinger@ethz.ch).}

	\begin{abstract}                
		In large-scale networks of uncertain dynamical systems, where communication is limited and there is a strong interaction among subsystems, learning local models and control policies offers great potential for designing high-performance controllers. At the same time, the lack of safety guarantees, here considered in the form of constraint satisfaction, prevents the use of data-driven techniques to safety-critical distributed systems. This paper presents a safety framework that guarantees constraint satisfaction for uncertain distributed systems while learning. The framework considers linear systems with coupling in the dynamics and subject to bounded parametric uncertainty, and makes use of robust invariance to guarantee safety. In particular, a robust non-convex invariant set, given by the union of multiple ellipsoidal invariant sets, and a nonlinear backup control law, given by the combination of multiple stabilizing linear feedbacks, are computed offline. In presence of unsafe inputs, the safety framework applies the backup control law, preventing the system to violate the constraints. As the robust invariant set and the backup stabilizing controller are computed offline, the online operations reduce to simple function evaluations, which enables the use of the proposed framework on systems with limited computational resources. The capabilities of the safety framework are illustrated by three numerical examples.
	\end{abstract}


\end{frontmatter}

\copyrightnotice
\vspace{-0.4cm}

\section{Introduction}
Large-scale safety-critical distributed systems are challenging to model and control due to their complexity, the interactions between subsystems, and their limited communication capabilities. Examples include power networks, automated warehouses, or vehicle platoons. Data-driven approaches for learning and control are appealing to overcome these challenges, and have shown superior performance compared to traditional control design methodologies via their capability of adapting the model and controller from data, see e.g.~\cite{Busoniu2008,Tuyls12,Nguyen18}. However, despite the promising performance shown in research-driven applications~\cite{Kar2013b,Kar2013a,Varshavskaya2009}, and~\cite{Zhang2018}, the applicability of data-driven techniques to safety-critical systems is severely limited by the lack of theoretical guarantees on the system behavior. Using learning-based controllers in closed-loop may lead the system to unwanted and dangerous behavior, e.g., in the form of constraint violation and instability. While safe learning algorithms for single systems have been proposed, e.g. in~\cite{Gillula11,Berkenkamp15,Wabersich2018b,Wabersich2019,Fisac19}, the problem of guaranteeing constraint satisfaction while learning in multi-agent systems has received less attention, see e.g.~\cite{Larsen2017},~\cite{Muntwiler2019}, and~\cite{Chen2019}.
This paper introduces a distributed framework for safe learning, in the form of an algorithm that prevents constraint violation at all times based on local conditions and controllers. We consider constrained distributed linear systems subject to bounded parametric uncertainties, and employ structured robust invariance at the core of the framework. The set of safe states is defined as the union of multiple ellipsoidal robust invariant sets, where each ellipsoidal robust invariant set is associated with a stabilizing linear feedback law. An input is considered unsafe if the one-step prediction leads the system state outside of the safe region. In the presence of unsafe inputs, the safety framework applies a combination of backup control laws to prevent the system from violating state and input constraints. Both, ellipsoidal sets and linear feedback laws are jointly precomputed offline via a distributed convex optimization problem. This reduces the online operation to a simple function evaluation, rendering the approach, feasible for real-time application.

\textit{Related Work:} In recent years, the interest in safe learning has grown, see e.g.~\cite{Garcia2015a} for a comprehensive overview in the context of reinforcement learning. Reachability analysis for dynamic systems is adopted in~\cite{Gillula11} and~\cite{Gillula12} to construct a safe set and backup controller. Safety frameworks based on Model Predictive Control (MPC) have been presented in~\cite{Wabersich2018a,Wabersich2018b,Wabersich2019}, and~\cite{Koller2018}, where robust and stochastic MPC formulations are used to guarantee constraint satisfaction at all times.

Safety frameworks for distributed linear systems with additive uncertainty were introduced in~\cite{Larsen2017} and~\cite{Muntwiler2019}. The work in~\cite{Larsen2017} is based on the computation of structured ellipsoidal robust invariant sets.  The approach presented in~\cite{Muntwiler2019} is instead based on a distributed MPC formulation, which is especially beneficial if the optimal operation point of the network is close to the border of the state constraints, but at the cost of a higher computational burden. A multi-agent Lyapunov-based safe feedback is also presented in~\cite{Wieland2007}, and the proposed scheme is applied to a collision avoidance problem. The safety framework in~\cite{Chen2019} exploits control barrier functions and robust invariant sets to guarantee constraint satisfaction, where the dynamic interaction among subsystems is reduced using assume-guarantee contracts.

\textit{Contributions:}
The contribution of this paper is twofold. First, we propose a safety framework that makes use of robust invariant sets and stabilizing linear feedback controllers for distributed constrained linear systems with parametric uncertainty. The computation of the invariant sets is based on the work presented in~\cite{Conte2016}. Differently from the distributed safety framework introduced in~\cite{Larsen2017}, we consider parametric uncertainty instead of additive uncertainty, and make use of multiple ellipsoidal robust invariant sets, thereby reducing conservatism. Moreover, the optimization problem for computing the invariant sets is a linear semidefinite program (SDP) instead of a bilinear SDP. This renders the optimization problem free of tuning parameters and allows the use of standard SDP solvers, such as MOSEK or SDPT3, see~\cite{mosek} and \cite{Tutuncu2003}, respectively. The second contribution is an explicit description of the safe set through the union of ellipsoidal invariant sets. Differently from~\cite{Muntwiler2019}, the proposed algorithm moves most of the computational burden offline, and the online operations reduce to simple function evaluations, rendering the approach feasible for real-time application.

The remainder of the paper is organized as follows: Section~\ref{sec:preliminaries} introduces preliminaries and the problem formulation. The distributed safety framework is presented in Section~\ref{sec:dsf}, and Section~\ref{sec:numerical_results} discusses the numerical results. Finally, Section~\ref{sec:conclusions} concludes the paper.

\section{Preliminaries}\label{sec:preliminaries}
\subsection{Notation}
Let~$\col(x_i)$ define a column vector obtained by stacking the vectors $x_i$. The symbol $\geq$ ($\leq$) used with matrices indicates a positive (negative) semi-definite matrix. The notation $[ \cdot ]_l$ applied to a vector or matrix, denotes the l-th row of the vector or matrix. Given a matrix $A \in \mathbb{R}^n$, the trace of $A$ is indicated with $\text{tr}(A)$.

\subsection{Problem Formulation}
\label{sec:problem_formulation}
Let $\G = (\V,\E)$ be an undirected communication graph, where $\V = \{1,\dots,N\}$ represents the nodes of the network, and $\E \subseteq \V \times \V$ contains the unordered pairs of nodes~$\{i,j\}$ which can communicate with each other. We denote with $\Ni =\{j \mid {i,j} \in \E\} \subseteq \V$ the set of neighbors of node $i$. We associate with each node $i \in \V$ a discrete-time dynamical linear system with parametric uncertainty
\begin{equation}\label{eq:agent_model}
	x_{i,k+1} = A_i\left(\theta_i\right) x_{\Ni ,k} + B_i\left(\theta_i\right) u_{i,k},
\end{equation}
where $x_i \in \mathbb{R}^{n_i}$, $x_{\Ni} \in \mathbb{R}^{n_{\Ni}}$, $u_i \in \mathbb{R}^{m_i}$ are the state of agent $i$, the state of agent $i$ and its neighbors, and the input of agent $i$, respectively. The uncertain vector~$\theta_i$ belongs to a polytopic set $\Theta_i \subset \mathbb{R}^{p_i}$, i.e. $\theta_i\in\Theta_i$.
Each subsystem~\eqref{eq:agent_model} is subject to polytopic state and input constraints containing the origin in their interior
\begin{equation}\label{eq:agent_constraints}
	\begin{split}
		&\mathbb{X}_i = \{x_{\Ni} \mid H_i x_{\Ni} \leq h_{i}  \}, \\
		&\mathbb{U}_i = \{u_{i} \mid  O_i u_{i}    \leq o_{i} \},
	\end{split}
\end{equation}
where $H_i \in \mathbb{R}^{n_{h_i} \times {n_{\Ni}}}, h_i \in \mathbb{R}^{n_{h_i}}, O_i \in \mathbb{R}^{n_{o_i} \times m_i},$ and $o_i \in \mathbb{R}^{n_{o_i}}$. Note that the state constraints can couple neighboring subsystems, while the input constraints only act locally.
The dynamics of the overall network is given by
\begin{equation}\label{eq:global_dynamics}
	x_{k+1} = A(\theta)x_k + B(\theta)u_k,
\end{equation}
where $x = \col(x_i)$, $u=\col(u_i)$, and $\theta=\col(\theta_i)$ represent the state, the input, and the uncertain parameter vector, respectively. The matrix $A \in \mathbb{R}^{n \times n}$ is an appropriate composition of the blocks $A_i$, and the input matrix is block diagonal, i.e., $B = \diag(B_1,\ldots,B_N)$. The state constraints for system~\eqref{eq:global_dynamics} are given by  $\mathbb{X} = \{x \mid x_{\Ni} \in \mathbb{X}_i\  \forall i \in \V\}$. Taking the cartesian product of all local input constraints, we denote the global input constraint set as $\mathbb{U} = \mathbb{U}_1 \times \cdots \times \mathbb{U}_N$.
Additionally, we denote with $T_i \in \left\lbrace 0,1\right\rbrace^{n_i \times n}$ and $W_i \in \left\lbrace 0,1\right\rbrace^{n_{\mathcal{N}_i} \times n}$ two lifting matrices such that $x_i = T_ix$ and $x_{\mathcal{N}_i} = W_ix$.

In the context of this paper, we refer to a learning algorithm as a generic procedure used to either retrieve information from data during closed-loop operation, e.g., to learn a model, see e.g.~\cite{Chiuso2019}, or to train learning-based controllers, e.g., via Q-learning,~\cite{Watkins1992}, or Proximal Policy Optimization (PPO),~~\cite{Schulman2017}. The learning algorithm would apply an input $u_i^L$ to each subsystem in order to achieve its goal. Most available learning algorithms, however, do not guarantee the satisfaction of state and input constraints. Constraint satisfaction is particularly challenging and crucial in distributed safety-critical applications, where the interactions among subsystems can render the global system unstable, see~\cite{Blanchini2015}.

In this paper, we propose a safety framework for distributed systems that can augment any learning algorithm to guarantee satisfaction of state and input constraints. The proposed safety framework verifies if an input is safe to apply and, if it is not, provides a backup control law that guarantees constraint satisfaction for all future times. Formally this is captured by the following definition of a safety framework, a similar definition was introduced in~\cite{Wabersich2018b}.
\begin{definition}
	\label{def:safety_framework}
	A distributed safety framework, defined by the functions $g_i(x_{\Ni},u_i):\mathbb{R}^{n_{\Ni}} \times \mathbb{R}^{m_i} \rightarrow \mathbb{R}^{m_i} \ $ for $ i \in \V$, certifies an input $u_{i,\bar k}^L$ as safe for system~\eqref{eq:agent_model} at iteration $\bar k$, if $u_{i,\bar k}^L = g_i(x_{\Ni,\bar k},u_{i,\bar k}^L)$. The application of $u_{i,k} = g_i(x_{\Ni, k},u_{i,k}^L)$ for all $i \in \V$ and for $k \geq \bar k$ guarantees the satisfaction of constraints~\eqref{eq:agent_constraints} at all times.
\end{definition}

\section{Distributed Invariance-based Safety Framework}
\label{sec:dsf}
In this section, we present the proposed distributed safety framework for constrained linear systems subject to bounded parametric uncertainty, defined in~\eqref{eq:agent_model}. We address the problem by first introducing a safety framework that implicitly defines the safe states and computes an input that keeps the system safe by solving an optimization problem. In order to address real-time requirements, we then derive an explicit approximation of the safety framework. A safe region of the state space and a backup control law are explicitly computed in the form of a union of robust invariant sets and stabilizing linear feedback laws.

\subsection{Implicit Formulation}
\label{subsec:dsf}

The safety framework is based on the concepts of robust invariance and structured invariant sets, proposed in~\cite{Conte2016}. The algorithm has two key features: 1) it provides an implicit description of the set of safe states and safe inputs through a convex optimization problem, 2) it allows for distributed computation.

The safety framework works as follows: Given a state $x_{i,k}\ \forall i \in \V$, we certify a learning input~$u_{i,k}^L$ as safe, by verifying that the uncertain one-step prediction lies within
a robust invariant set of the form $\mathcal{X} = \{ x \mid x^TPx = \sum_{i=1}^N x_i^TP_ix_i \leq 1 \} \subseteq \mathbb{X}$ for system~\eqref{eq:agent_model} under a stabilizing control law~$u_{i,k} = K_i x_{\Ni}$, where $K_i \in \mathbb{R}^{m_i \times n_{\Ni}}$, and $P = \diag(P_1, \ldots, P_N)$.  The following conditions are sufficient for the existence of such a set:
\begin{align}
	  & (A(\theta) + B(\theta)K)^T P (A(\theta) + B(\theta)K) -P \leq 0, \label{eq:lyap_decrease}                   \\
	  & \sum_{i=1}^N x_{i,k+1}^TP_ix_{i,k+1} \leq 1, \label{eq:one_step_prediction_containment}                     \\
	  & P_i > 0\quad \forall i \in \V, \label{eq:local_psd}                                                         \\
	  & x_{i,k+1} = A_i\left(\theta_i\right) x_{\Ni ,k} + B_i\left(\theta_i\right) u^L_{i,k}, \label{eq:prediction}
\end{align}
for all possible uncertain parameter values $\theta_i$, where $K$ is an appropriate composition of the structured linear feedback gains $K_i$. Note that~\eqref{eq:lyap_decrease} requires a matrix $P$ which makes $\mathcal{X}$ a robust ellipsoidal invariant set for the uncertain autonomous linear system $A(\theta) + B(\theta)K$. The LMI~\eqref{eq:one_step_prediction_containment} verifies whether the learning inputs $u_{i,k}^L$ are safe, i.e., the one-step predictions satisfy $x_{i,k+1} \in \mathcal{X}_i \ \forall i \in \V$.
\begin{remark}
	Condition~\eqref{eq:lyap_decrease} is a global condition, i.e., it has to hold for the whole network of systems. However, this condition can be distributed by upper bounding the global Lyapunov decrease with a structured symmetric matrix. More details are given in the proof of Proposition~\ref{prop:implicit_dsf} and can be found in~\cite{Conte2016}.
\end{remark}
In order to define a distributed safety framework that interferes as little as possible with the learning algorithm, we propose the following approach. We introduce an optimization problem that ensures feasibility of~\eqref{eq:lyap_decrease} -~\eqref{eq:prediction} for a safe input that is as close as possible to the learning input~$u_{i,k}^L$. This can be achieved by defining the safe input as $u_{i,k} = u_{i,k}^L + \Delta u_{i,k}$, and minimizing~$\Delta u_{i,k}$ subject to~\eqref{eq:lyap_decrease} -~\eqref{eq:prediction}, and $u_{i,k} \in \mathbb{U}_i$.

The following proposition formalizes the distributed safety framework described above.
\begin{proposition}{}\label{prop:implicit_dsf}
	Consider system~\eqref{eq:agent_model} subject to~\eqref{eq:agent_constraints}. Assume $x_{i,0}\ \forall i \in \V$ to be a feasible initial condition for the following optimization problem
	\begin{subequations}\label{eq:distributed_sf}
		\begin{alignat}{2}
			\min_{E,Y,S,\Delta u} \quad & \norm{\Delta u_{i,k}}^2                                                                                                         \\
			\textrm{s.t. }              & \forall l \in \left\lbrace 1,\dots ,n_{h_i}\right\rbrace,\  \forall e \in \left\lbrace 1,\dots ,n_{o_i}\right\rbrace, \nonumber \\
			                            & \forall i \in \V,\  \theta_i \in \Theta_i,                                                                                      \\
			                            & E_i > 0,       \label{eq:e_positive_definite}                                                                                   \\
			                            & \begin{bmatrix}\overline{E}_i + \mathcal{S}_{\mathcal{N}_i} & \star\\ A_{\Ni}(\theta_i)E_{\mathcal{N}_i}+B_i(\theta_i)Y_i & E_i\end{bmatrix} \geq 0,   \label{eq:invariance}                                                                       \\
			                            & \sum_{r \in \mathcal{N}_i} T_iW_r^TS_{\mathcal{N}_r}W_rT_i^T \leq 0, \label{eq:distributed_invariance}                          \\
			                            & \begin{bmatrix} \nicefrac{1}{N} & x_{i,k+1}^T\\ x_{i,k+1} & E_i \end{bmatrix} \geq 0,   \label{eq:robust_prediction}                                                               \\
			                            & O_{i}\left( \Delta u_{i,k} + u_{i,k}^L \right) \leq o_{i}, \label{eq:safe_input}                                                \\
			                            & \begin{bmatrix} \left[h_{i}\right]_l^2 & \left[H_{i}\right]_{l,*}E_i\\ E_i\left[H_{i}\right]_{l,*}^T & E_i \end{bmatrix} \geq 0,   \label{eq:state_constraints}                                                               \\
			                            & \begin{bmatrix} \left[o_{i}\right]_e^2 & \left[O_{i}\right]_{e,*}Y_i\\ Y_i^T\left[O_{i}\right]_{e,*}^T & E_{\mathcal{N}_i} \end{bmatrix} \geq 0, \label{eq:input_constraints}                                                                 \\
			                            & x_{i,k+1} = A_{\mathcal{N}_i}(\theta_i)x_{\mathcal{N}_i} + B_i(\theta_i)(u_{i,k}^L + \Delta u_{i,k} )
		\end{alignat}
	\end{subequations}
	where $E_i=P_i^{-1}$, $E_{\Ni} = W_iEW_i^T$, $\overline{E}_i = W_iT_i^TE_iT_iW_i^T$, $\mathcal{S}_{\mathcal{N}_i} = E_{\Ni}\Gamma_{\mathcal{N}_i}E_{\Ni} $  and $Y_i = K_iE_{\Ni}$. Then
	\begin{equation} \label{eq:implicit_sf}
		g_i(x_{\Ni,k},u_{i,k}^L) = u_{i,k}^L + \Delta u_{i,k}
	\end{equation}
	defines a distributed safety framework according to Definition~\ref{def:safety_framework}, where $\Delta u_{i,k}$ is the solution of~\eqref{eq:distributed_sf}.
\end{proposition}
\begin{pf}
	The proof is structured in three parts proving that $i)$ inequalities~\eqref{eq:e_positive_definite}-\eqref{eq:distributed_invariance} define a structured robust invariant set $\mathcal{X}$, $ii)$ inequalities~\eqref{eq:robust_prediction}-\eqref{eq:safe_input} determine a safe input $u_k$, and $iii)$ inequalities~\eqref{eq:state_constraints}-\eqref{eq:input_constraints} ensure the feasibility of the states and the inputs.
	\begin{enumerate}[i)]
		\item Using the distribution of the Lyapunov decrease condition proposed in~\cite{Conte2016}, we reformulate~\eqref{eq:lyap_decrease} as a set of local inequalities coupled by a global condition
		      \begin{align}
			        & (A_{\Ni}(\theta_i) + B_i(\theta_i)K_i)^T P_i (A_{\Ni}(\theta_i) + B_i(\theta_i)K_i) \nonumber \\
			        & - W_iT_i^TP_iT_iW_i^T \leq \Gamma_{\Ni} \quad \forall i \in \V, \label{eq:local_inequalities} \\
			        & \sum_{r = 0}^N W_r^T\Gamma_{\mathcal{N}_i}W_r \leq 0. \label{eq:global_coupling_condition}
		      \end{align}
		      Using the Schur complement on~\eqref{eq:local_inequalities} and relaxing condition \eqref{eq:global_coupling_condition} by upper bounding it with block diagnonal matrices $S_{\Ni}$, we obtain the LMIs~\eqref{eq:e_positive_definite}-\eqref{eq:distributed_invariance}.
		\item Using the Schur complement on~\eqref{eq:one_step_prediction_containment} we obtain
		      $$
			      \begin{bmatrix} \nicefrac{1}{N} & x_{i,k+1}^T\\ x_{i,k+1} & E_i \end{bmatrix} \geq 0,\  \forall i \in \V,
		      $$
		      where we use the fact that $P$ and therefore $E = P^{-1}$ is block-diagonal. Condition~\eqref{eq:robust_prediction} guarantees that the uncertain one-step prediction lies within the structured robust invariant set $\mathcal{X}_i$, while~\eqref{eq:safe_input} verifies that $u_{i,k}$ satisfies the input constraints.
		\item Last, we consider the feasibility of the states and inputs, which is guaranteed if $$\mathcal{X} \subseteq \left\lbrace x \mid O_i K_i x_{\mathcal{N}_i} \leq o_i, \;\forall i \in \V \right\rbrace \cap \mathbb{X}.$$ Using standard techniques for containment of ellipsoids in polyhedra, see e.g.~\cite{Boyd:94}, this can be expressed as~\eqref{eq:state_constraints} and~\eqref{eq:input_constraints}, see e.g.~\cite{Gao01}.\hfill\ensuremath{\square}
	\end{enumerate}
\end{pf}
Note that the objective of~\eqref{eq:distributed_sf} is to minimize the modification of the learning inputs $u_{i,k}^L$ given by $\Delta u_{i,k}$. Problem~\eqref{eq:distributed_sf} is convex and structured, which renders the optimization problem amenable to distributed optimization, see e.g.~\cite{bertsekas89} for an overview. Checking all possible realizations of the uncertain parameter space $\Theta_i$ could be computationally intractable. However, given the fact that we are considering linear systems, it is sufficient to check the extremal points of the polytopic sets~$\Theta_i$ to compute~\eqref{eq:distributed_sf} $\forall \theta_i \in \Theta_i$, see e.g.~\cite{Kharitonov78}.

The presented safety framework requires to solve~\eqref{eq:distributed_sf} at every iteration, which can be computationally demanding. In the following section, we introduce an approximation that reduces the online computational burden to simple function evaluations.

\subsection{Explicit Approximation}
\label{subsec:explicit_dsf}

The explicit safety framework derived in this section takes inspiration from explicit model predictive control (MPC),~\cite{Bemporad2002}, and locally weighted projection regression (LWPR),~\cite{vijayakumar2000}. Explicit MPC removes the need to solve an optimization problem online by pre-computing the control law offline, such that the online operations reduce to simple function evaluations. LWPR achieves nonlinear function approximation in high dimensional spaces by partitioning the feature space using data, and fitting local linear models to each partition.

\begin{figure}[t]
	\centering
	\includegraphics{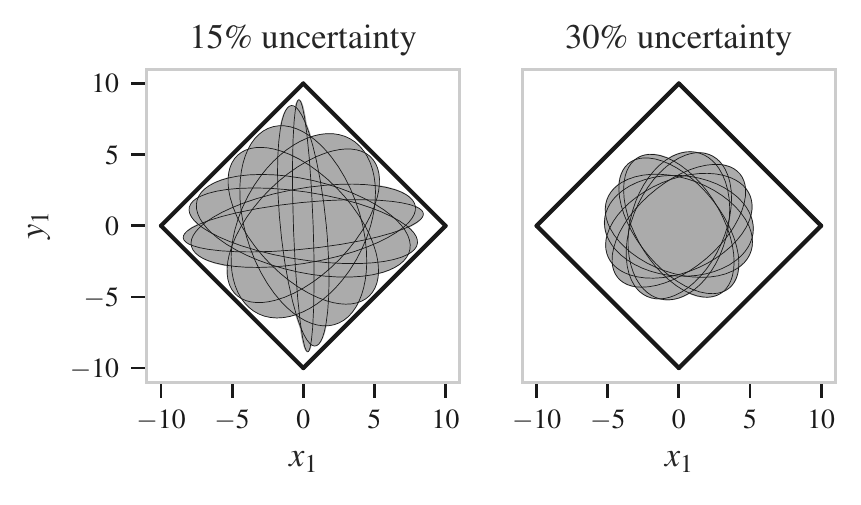}
	\caption{Invariant sets for one agent, using 10 partitions with an uncertainty level of 15\% and 30\%.}
	\label{fig:unionPlots}
\end{figure}

The proposed explicit safety framework consists of two phases: the initialization and the learning phase. The former is an offline phase, where the safe set is explicitly approximated. First, the state space is divided into $M$ regions $\mathcal{R}^j$ with $j \in \M = \{1,\ldots, M\},$ ensuring coverage of the whole state space. Subsequently, a structured ellipsoidal robust invariant set $\mathcal{X}^j =\{x \mid \sum_{i=1}^N x_i^T P_i^j x_i \leq 1 \} \subset \mathcal{X}$, such that $\mathcal{X}^j \cap \mathcal{R}^j \neq \emptyset$, and a corresponding structured linear stabilizing controller $K_{i}^j \in \mathbb{R}^{m_i \times n_{\Ni}}, \ \forall i \in \V$, is computed for every region~$\mathcal{R}^j$. The overall safe set is defined as the union of all robust invariant sets~$\mathcal{X}^j$. By partitioning the state space and requiring non-empty intersection of each partition and the corresponding robust invariant set, as well as by computing a linear feedback law for each invariant set, the resulting robust invariant sets will have different shapes and orientations, see e.g. Figure~\ref{fig:unionPlots}. The union of ellipsoidal robust invariant sets will thereby result in a larger safe set than achievable with a single robust ellipsoidal invariant set.

The state space, delimited by the state constraints~\eqref{eq:agent_constraints}, is divided into polytopic regions of the form $\mathcal{R}^j = \{\ x \mid A_{\mathcal{R}^j} x \leq b_{\mathcal{R}^j} \}$ using Voronoi partitioning, see e.g.~\cite{Cortes2004}. The structured robust invariant sets $\mathcal{X}^j$ are computed in a distributed manner by modifying~\eqref{eq:distributed_sf}. The non-empty intersection between the robust invariant set~$\mathcal{X}^j$ and the region $\mathcal{R}^j$ is enforced with the following convex constraint
\begin{subequations}
	\begin{align}
		  & x^TP^jx \leq 1,  \label{eq:safe_set}                                 \\
		  & A_{\mathcal{R}^j} x \leq b_{\mathcal{R}^j} \label{eq:R_containment}.
	\end{align}
\end{subequations}
The computation of the structured invariant sets and the stabilizing control law is formalized by the following proposition.
\begin{proposition}{}\label{prop:explicit_offline_dsf}
	Consider system~\eqref{eq:agent_model} subject to~\eqref{eq:agent_constraints}. Given $M$ convex regions of the form $\mathcal{R}^j = \{\ x \mid A_{\mathcal{R}^j} x \leq b_{\mathcal{R}^j} \}$, $M$ structured robust invariant sets $\mathcal{X}^j=\{x \mid \sum_{i=1}^N x_i^T P_i^j x_i \leq 1\}$ such that $\mathcal{X}^j \cap \mathcal{R}^j \neq \emptyset$, and $M$ structured linear stabilizing control laws $u_{i}^j = K_{i}^j x_{\Ni}\  \forall i \in\V$ with $j \in \M$ can be computed with the following optimization problem:
	\begin{subequations}\label{eq:offline_explicit_dsf}
		\begin{align}
			\min_{E,Y,S,x} \quad & \sum_{i=1}^N\textrm{tr}  \left(E_i^j\right)                                                                                      \\
			\textrm{s.t.} \quad  & \forall l \in \left\lbrace 1,\dots ,n_{h_i}\right\rbrace,\  \forall e \in \left\lbrace 1,\dots ,n_{o_i}\right\rbrace,  \nonumber \\
			                     & \forall i \in \V, \ \forall j \in \{1,\ldots,M\},\  \theta_i \in \Theta_i,                                                       \\
			                     & E_i^j > 0, \label{eq:explicit_offline_pos_def}                                                                                   \\
			                     & \begin{bmatrix}\overline{E}_i^j + \mathcal{S}_{\mathcal{N}_i}^j & \star\\ A_{\Ni}(\theta_i)E_{\mathcal{N}_i}^j+B_i(\theta_i)Y_i^j & E_i^j\end{bmatrix} \geq 0,                                                                                               \\
			                     & \sum_{r \in \mathcal{N}_i} T_iW_r^TS_{\mathcal{N}_r}^jW_rT_i^T \leq 0,                                                           \\
			                     & \begin{bmatrix} \left[h_i\right]_l^2 & \left[H_i\right]_{l,*}E_{\Ni}^j\\ E_{\Ni}^j\left[H_i\right]_{l,*}^T & E_{\Ni}^j \end{bmatrix} \geq 0,                                                                                               \\
			                     & \begin{bmatrix} \left[o_i\right]_e^2 & \left[O_i\right]_{e,*}Y_i^j\\ (Y_i^j)^T\left[O_i\right]_{e,*}^T & E_{\Ni}^j \end{bmatrix} \geq 0,      \label{eq:explicit_offline_input_constraints}                                            \\
			                     & \begin{bmatrix} \nicefrac{1}{N} & x_i^T\\ x_i & E_i^j \end{bmatrix} \geq 0, \label{eq:coverofR}                                                                           \\
			                     & W_i A_{\mathcal{R}^j} W_i^T x_{\Ni} \leq W_i b_{\mathcal{R}^j}, \label{eq:Rset}
		\end{align}
	\end{subequations}
	where $E_i^j=(P_i^j)^{-1}$, $E_{\Ni}^j = W_i^TE^jW_i$ , $\overline{E}_i^j = W_iT_i^TE_i^jT_iW_i^T$, and $Y_i^j = K_i^jE_{\Ni}^j$.
\end{proposition}

Problem~\eqref{eq:offline_explicit_dsf} is convex and can be solved in a distributed manner using the same techniques as described in Subsection~\ref{subsec:dsf}.

We now describe the learning phase. Given the state $x_{i,k}\ \forall i \in \V$, the safety framework applies the learning input~$u^L_{i,k}\ \forall i \in \V$ if the uncertain one-step prediction lies within one of the robust invariant sets $\mathcal{X}^j$ with $j \in \M$. Otherwise, the safety framework applies one of the precomputed structured linear stabilizing controllers $u_{i,k} = K_{i}^{b^\star} x_{\Ni,k}\  \forall i \in \V$, where $b^\star$ is the index of the robust invariant set that contains the current state and results in the smallest modification of the learning input, i.e.,
\begin{equation}
	\label{eq:best_safe_controller}
	\begin{split}
		b^\star = \argmin_{b} &\sum_{i=1}^N \norm{u_{i,k}^L - K_i^b x_{\Ni,k}} \\
		&\text{s.t. } x_{i,k} \in \mathcal{X}^b_i, \ \forall i \in \V.
	\end{split}
\end{equation}
We formalize the safety framework outlined above in the following proposition.

\begin{proposition}{}\label{prop:explicit_dsf}
	Consider system~\eqref{eq:agent_model} subject to~\eqref{eq:agent_constraints}, given the initial condition $x_{i,0} \in \mathcal{X}_i^j\  \forall i \in \V$ for some $j \in \M$, where $\mathcal{X}_i^j$ are computed using~\eqref{eq:offline_explicit_dsf}. Then
	\begin{equation}\label{eq:explicit_control_law}
		g_i(x_{\Ni,k},u_{i,k}^L) =
		\begin{cases}
			\begin{matrix} u_{i,k}^L\ \  \textrm{ if }\  \exists j\in \M :\; x_{i,k+1} \in \mathcal{X}_i^j, \forall i \in \V \\
				K^{b^\star}_i x_{\Ni,k}\ \   \textrm{ otherwise, } \hfill
			\end{matrix} \
		\end{cases}
	\end{equation}
	$\forall i \in \V$, with $b^\star$ in~\eqref{eq:best_safe_controller}, defines a distributed safety framework according to Definition~\ref{def:safety_framework}.
\end{proposition}
\begin{pf}
	We prove the two distinct cases in~\eqref{eq:explicit_control_law} separately. The first case follows directly from the fact that $\mathcal{X}^j \subset \mathbb{X}$ and constraint satisfaction in the future follows from robust invariance of $\mathcal{X}^j$. The second case is trivial since $K^{b^\star}_i$ is one of the safe control laws $K_i^j$, thus $K^{b^\star}_i x_{\Ni,k}$ is always a safe input. Therefore, \eqref{eq:explicit_control_law} defines a safety framework according to Definition~\ref{def:safety_framework}. \hfill\ensuremath{\square}
\end{pf}

Note that the safety framework defined by~\eqref{eq:explicit_control_law} can be computed in a distributed fashion. The containment in the invariant set can be cast as an averaging consensus problem, where the cost to average is
$$
	J(x) = \frac{1}{N}x^TP^jx = \frac{1}{N}\sum_{i=1}^N x_i^T P_i^jx_i \ \forall j \in \M.
$$
The containment is guaranteed if and only if the consensus algorithm converges to a steady state smaller than or equal to~$\frac{1}{N}$, for more information on consensus problems see~\cite{Olfati-Saber2007}. Similarly, the optimization problem~\eqref{eq:best_safe_controller} can be solved using min-consensus where the containment in the invariant set can be verified as before. The explicit formulation of the safety framework is summarized in Algorithm~\ref{alg:algorithm_explicit_dsf}.

\begin{algorithm}[ht]
	\caption{Explicit Distributed Invariance-based Safety Framework}\label{alg:algorithm_explicit_dsf}
	\begin{algorithmic}[1]
		\STATE \textbf{Intialization Phase} (offline)
		\STATE Compute the partitions $\mathcal{R}^j$
		\STATE Solve optimization problem~\eqref{eq:offline_explicit_dsf}
		\STATE
		\STATE \textbf{Learning Phase} (online)
		\FOR{$k=1,2,\ldots$}
		\FOR{$\forall i \in \V$}
		\STATE Measure $x_{i,k}$
		\STATE Exchange $x_{i,k}$ with neighbors
		\STATE Obtain learning input $u^L_{i,k}$
		\STATE Apply $u_{i,k}=g_i(x_{\Ni,k},u_{i,k}^L)$, defined by~\eqref{eq:explicit_control_law}
		\ENDFOR
		\ENDFOR
	\end{algorithmic}
\end{algorithm}

\section{Numerical Results}\label{sec:numerical_results}
In this section, we show the effectiveness of the proposed safety framework with three numerical examples. First, we illustrate the application of the explicit approximation presented in Subsection~\ref{subsec:explicit_dsf} in combination with a learning algorithm on a small-scale system. Subsequently, we also show the application to a large-scale system and end the section showing the quality of the explicit approximation compared to the implicit formulation presented in Subsection~\ref{subsec:dsf}. All examples are implemented in Python using CVXPY and OpenAI Gym, see~\cite{cvxpy} and~\cite{Gym}, respectively, and solved using MOSEK, see~\cite{mosek}. The examples were run on a machine equipped with an Intel Core i9 (3.1~GHz) CPU and 32~GB of RAM.

\subsection{Explicit Approximation in Combination with a Learning Algorithm}
In the first example, we make use of Proximal Policy Optimization (PPO), see~\cite{Schulman2017}, to learn a control policy for a three-agent mass-damper system, where the individual agent's dynamics are given by
\begin{equation}
	\begin{split}\label{eq:dynamics_ex3}
		m_i \ddot{x}_i &= a_i\dot{x}_i + d_{ij}\left(\dot{x}_j - \dot{x}_i\right) + u_i, \\
		m_i \ddot{y}_i &= a_i\dot{y}_i + d_{ij}\left(\dot{y}_j - \dot{y}_i\right) + u_i
	\end{split} \qquad \forall j \in \Ni,
\end{equation}
where $d_{ij}$ is the damping coefficient, $a_i$ is a parameter damping the agent's local dynamics, and $x_i$ and $y_i$ are the lateral and longitudinal axes in the two dimensional plane, respectively. The parameter vector $\theta_i$ consists of the parameter $a_i$ and the parameters $d_{ij}$, which couple agent $i$ to all neighboring agents. The uncertainty set reads as
$$\Theta_i = \left[ \left( 1 - \gamma \right)\theta_i^{nominal}, \left( 1 + \gamma \right)\theta_i^{nominal} \right],$$ where $\gamma$ is the uncertainty level, and $\theta_i^{nominal}$ is the true parameter vector, which is also used in the simulation model. The agents' state constraints are of the form~\eqref{eq:agent_constraints} and only constrain the local positions, while the local velocities are unconstrained. As the communication graph $\G$ we use again a line graph and the numerical values of the example are given in Table \ref{tab:example3}.

\begin{figure}[t]
	\centering
	\includegraphics{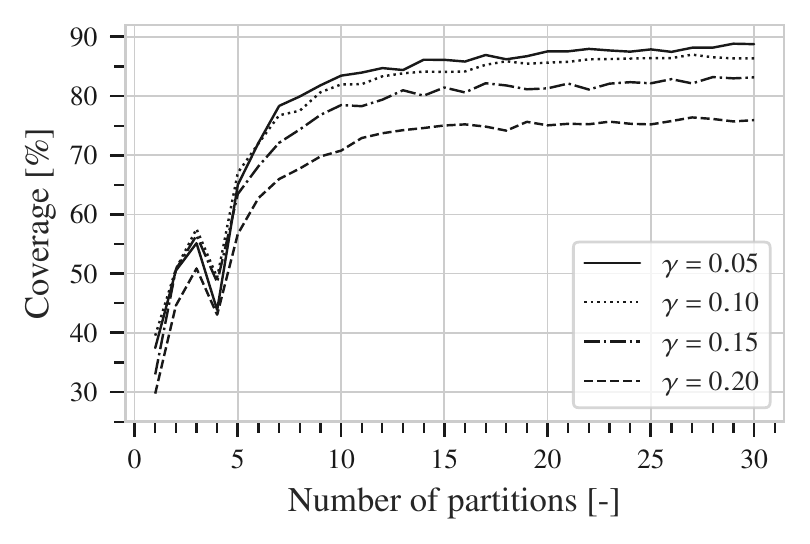}
	\caption{Fraction of the constrained state space covered by the union of pre-computed robust invariant sets for different numbers of partitions and different uncertainty levels.}
	\label{fig:coveragePlot}
\end{figure}

To measure the performance of the explicit approximation presented in Subsection~\ref{subsec:explicit_dsf}, we compute the fraction of the constrained state space covered by the union of pre-computed robust invariant sets, i.e., $\nicefrac{\text{vol}(\cup_{j=1}^N \mathcal{X}^j)}{\text{vol}(\cup_{i=1}^N \mathbb{X}_i)}$. Figure~\ref{fig:unionPlots} shows the union of the robust invariant sets for the first agent for two levels of uncertainty. The shown cross section illustrates the plane at zero velocity in both directions $x$ and $y$. The plot shows that a large portion of the state space is covered, and that the use of multiple robust invariant sets is beneficial to cover a larger area of the state space. Figure~\ref{fig:coveragePlot} shows the fraction of the constrained state space covered by the union of invariant sets as a function of the number of regions in a partition and the uncertainty level $\gamma$. Since the partitioning affects the computation of the robust invariant sets, the coverage results in Figure~\ref{fig:coveragePlot} are averaged over 20 Voronoi partitions obtained from random initial conditions.

\begin{figure}[t]
	\centering
	\includegraphics{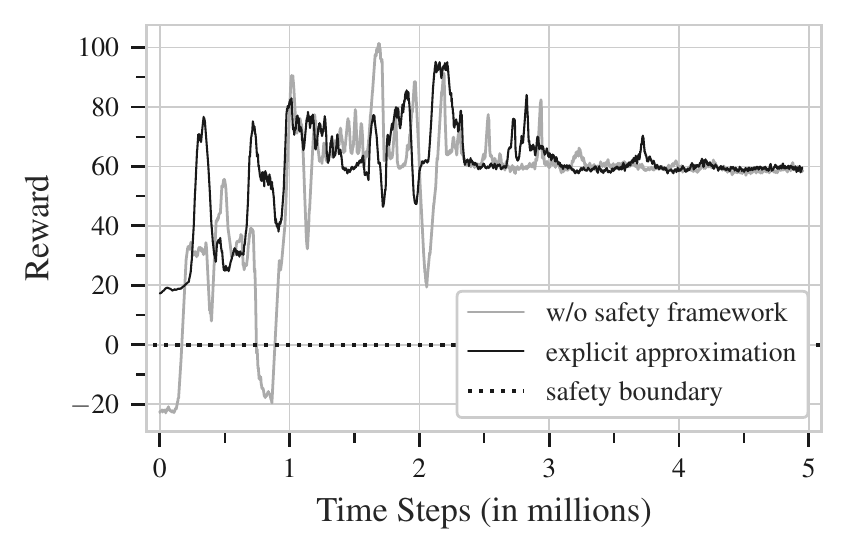}
	\caption{Learning curve for a simple regulation task, using PPO with and without the explicit approximation of the safety framework.}
	\label{fig:learningPlot}
\end{figure}

To demonstrate the ability of the explicit approximation to ensure constraint satisfaction, we consider the aforementioned system in a learning context, where the goal is to learn a controller, which regulates the state of each system back to the origin. We set the uncertainty level to 20\% and the number of partitions $M$ to 15. The learning setting is episodic, and at the start of each episode, each system is set to a random position in the two dimensional plane within the state constraints. For the learning algorithm we use PPO, which is also applicable in a multi-agent setting, see e.g.~\cite{Srinivasan2018}. The learning curve for PPO with and without the explicit approximation of the distributed safety framework is depicted in Figure~\ref{fig:learningPlot}. In this example, violating the state constraints would result in a negative reward, which never happens if the safety framework is active. In both cases, the learning algorithm converges in about the same number of iterations, therefore the safety framework has no negative affect on the learning rate in this example.

\begin{table}[h]
	\centering
	\caption{Numerical values for the first example.} \label{tab:example3}
	\begin{tabular}{ccccccc}
		$m_i$ & $a_i$ & $d_{ij}$ & $H_i$                        & $h_i$                        & $O_i$                        & $o_i$                        \\ \hline
		1     & 0.1   & 0.5      & $\begin{bmatrix} 10 & 0 & 10 & 0 \\ 10 & 0 & -10 & 0 \\ -10 & 0 & 10 & 0 \\ -10 & 0 & -10 & 0 \end{bmatrix}$ & $\begin{bmatrix} 100 \\ 100 \\ 100 \\ 100 \end{bmatrix}$ & $\begin{bmatrix} 1 \\ -1 \end{bmatrix}$ & $\begin{bmatrix} 5 \\ -5 \end{bmatrix}$
	\end{tabular}
\end{table}

\subsection {Guaranteed Safety on a Large-Scale System}
\begin{figure*}[h]
	\centering
	\includegraphics{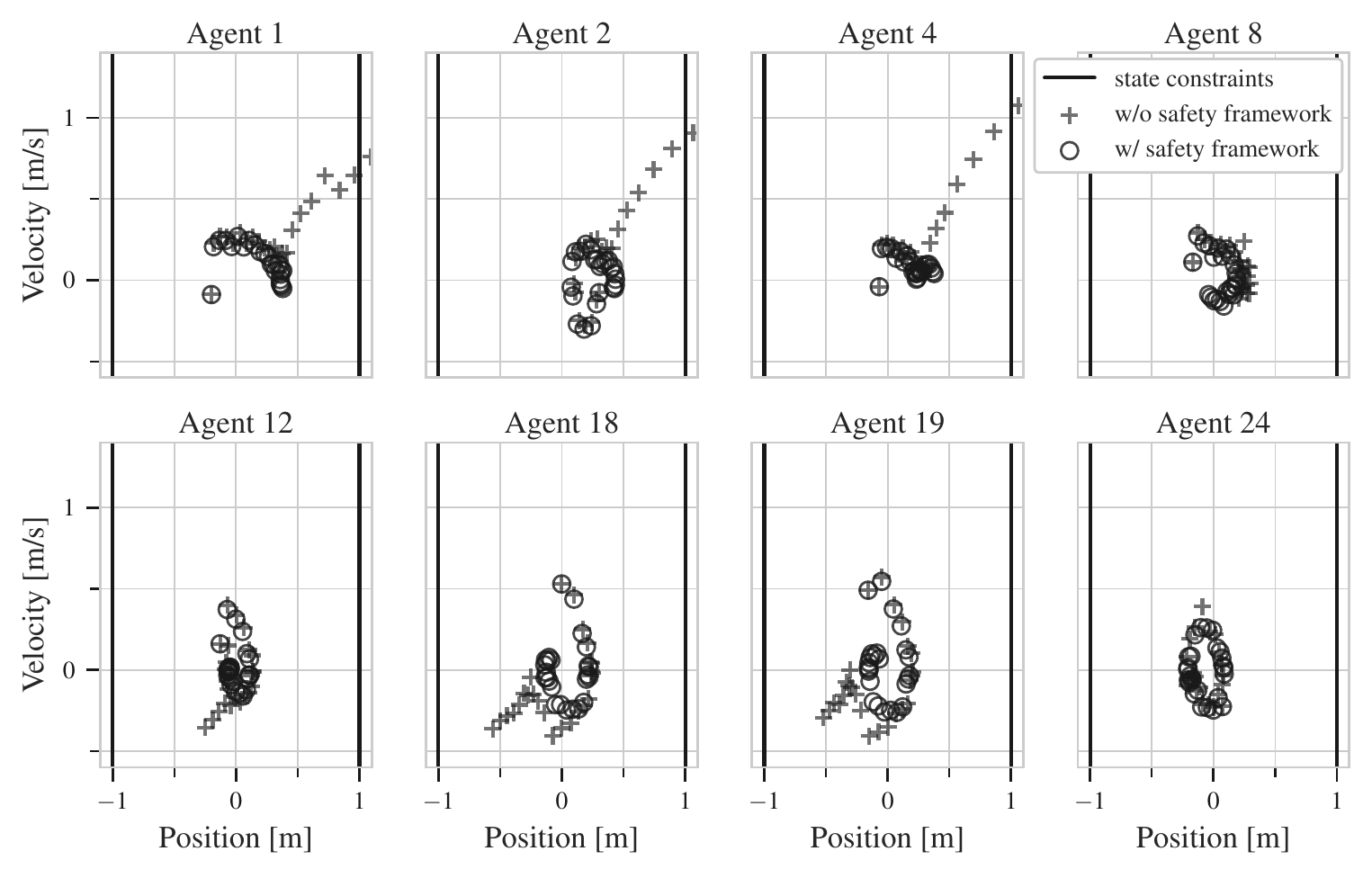}
	\caption{State evolution of eight selected agents driven by the unsafe exploration algorithm with and without the safety framework.}
	\label{fig:large-scale-example}
\end{figure*}

In the second example, we use a Gaussian Processes-based exploration algorithm, see e.g.~\cite{todescato2017multi}, in combination with the proposed safety framework to safely explore the state space of a large-scale system. The Gaussian Process is exploited to estimate the uncertainty at any point in the state space and to select the new input locations in order to reduce the maximum of the posterior covariance. We consider a 25-agent mass-spring-damper system, where the individual agent's dynamics are described by
\begin{equation}\label{eq:dynamics_ex1}
	m_i \ddot{x}_i = k_{ij}\left(x_j - x_i\right) + d_{ij}\left(\dot{x}_j - \dot{x}_i\right) + u_i \quad \forall j \in \Ni,
\end{equation}
where $k_{ij}$ and $d_{ij}$ are the spring and damping coefficients, respectively. Additionally, the dynamics~\eqref{eq:dynamics_ex1} are subject to constraints of the form~\eqref{eq:agent_constraints} and the communication graph $\G$ is a line graph. Table~\ref{tab:example1} provides the numerical values of both, the dynamical parameters and the constraint matrices.
Figure~\ref{fig:large-scale-example} shows the ability of the safety framework to keep the system safe, despite an input from the exploration algorithm, which drives some agents close to or beyond the state constraints, which is highlighted for a subset of the 25 agents. Note that the constraints shown in Figure~\ref{fig:large-scale-example} are the state constraints of the individual agents and do not include the additional constraints imposed by the dynamic coupling, i.e., the restrictions on the state of agent~$i$ due to the state of its neighbors, as these result in time-varying local constraints.

\begin{table}[thbp]
	\centering
	\caption{Numerical values for the second and third example.} \label{tab:example1}
	\begin{tabular}{ccccccc}
		$m_i$ & $k_{ij}$ & $d_{ij}$ & $H_i$                        & $h_i$                        & $O_i$                        & $o_i$                        \\ \hline
		1     & 2        & 1        & $\begin{bmatrix} 1 & 0 \\ -1 & 0 \\ 0 & 1 \\ 0 & -1 \end{bmatrix}$ & $\begin{bmatrix} 1 \\ -1 \\ 3 \\ -3 \end{bmatrix}$ & $\begin{bmatrix} 1 \\ -1 \end{bmatrix}$ & $\begin{bmatrix} 1 \\ -1 \end{bmatrix}$
	\end{tabular}
\end{table}

\subsection{Quality of the Explicit Approximation}
\begin{figure}[h]
	\centering
	\includegraphics{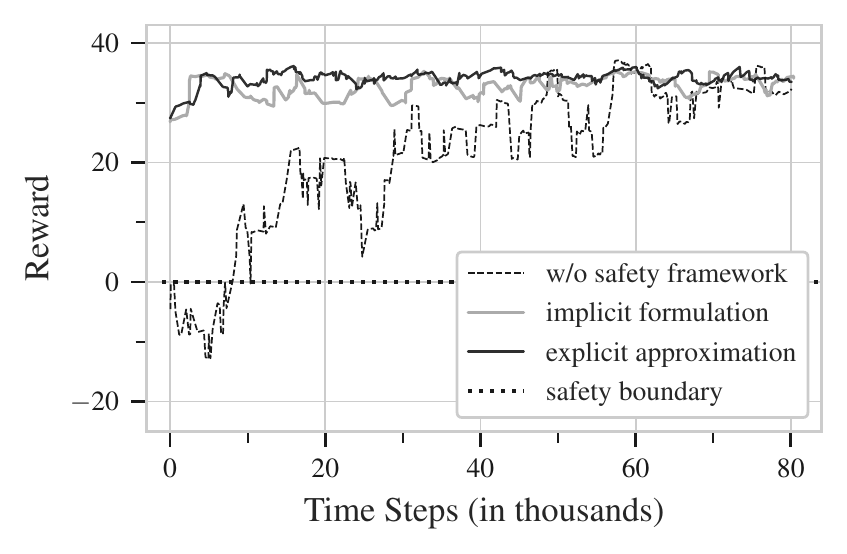}
	\caption{Learning curve for deep double Q-learning on a second-order system with three agents, using no safety framework, the implicit formulation of the safety framework, and its explicit approximation.}
	\label{fig:implicitvsexplicit}
\end{figure}

In the third example, we use deep double Q-learning, see~\cite{Hasselt15}, to learn a control policy that regulates a three-agent mass-spring-damper system back to the origin. The goal of this example is to evaluate the quality of the explicit approximation by comparing it against the implicit safety framework. We again use the model in~\eqref{eq:dynamics_ex1} with constraints~\eqref{eq:agent_constraints} as in the second example but reduce the number of agents, in order to facilitate the use of a reinforcement algorithm. Figure~\ref{fig:implicitvsexplicit} shows the learning curve in terms of the cumulative reward for the system without a safety framework, for the safety framework, and for its explicit approximation. We tuned the hyper-parameters of the Q-learning algorithm on the first setting and retrained it with the same hyper-parameters in the latter two settings. In all three settings, negative rewards indicate constraint violations. We can see from the figure that both, the distributed safety framework as well as the explicit approximation show comparable behavior and guarantee constraint satisfaction at all times.

\section{Conclusion}\label{sec:conclusions}
In this paper, we introduced an explicit safety framework, which guarantees constraint satisfaction for distributed uncertain linear systems based on local conditions. We showed with three numerical examples the effectiveness of the proposed algorithm in ensuring constraint satisfaction while learning.

\bibliography{bibliography}

\end{document}